\documentclass[preprint,12pt,authoryear]{elsarticle}

\usepackage{algorithm}
\usepackage{algpseudocode}
\usepackage{amssymb}
\usepackage{amsmath}
\usepackage{array}
\usepackage[hyphens]{url}
\usepackage{hyperref}
\usepackage{booktabs}
\usepackage{tabularx}
\usepackage{threeparttablex}
\usepackage{makecell}
\usepackage{tablefootnote}
\usepackage{pifont}
\usepackage{xcolor}

\newcommand{\cmark}{\makebox[1.2em][c]{\ding{51}}}
\newcommand{\xmark}{\makebox[1.2em][c]{\ding{55}}}
\newcolumntype{C}[1]{>{\centering\arraybackslash}m{#1}}
\newcolumntype{Y}{>{\raggedright\arraybackslash\hsize=2.5\hsize}X}
\newcolumntype{Z}{>{\centering\arraybackslash\hsize=0.7\hsize}X}

\journal{Energy Economics}

\begin{document}

\begin{frontmatter}



\title{Speculative Imbalance Positions by Non-Physical Traders: Impacts on Electricity System Balancing in Great Britain} %% Article title


\author[oxford]{Joseph Cary\corref{cor1}}
\cortext[cor1]{Corresponding author. Email address: \href{mailto:joseph.cary@wolfson.ox.ac.uk}{joseph.cary@wolfson.ox.ac.uk}}
\author[oxford]{Thomas Morstyn}
\affiliation[oxford]{organization={Department of Engineering Science, University of Oxford},
            addressline={Parks Road},
            city={Oxford},
            postcode={OX1 3PJ},
            state={Oxfordshire},
            country={United Kingdom}}
%% Abstract
\begin{abstract}
The electricity system in Great Britain (GB) allows participants to deliberately produce or consume a different volume to their traded position and has incentivised this practice with a single imbalance price since 2015. In particular, non-physical traders (NPTs) often take an open position to imbalance without any intention to produce or consume power. However, new government proposals could restrict this practice due to concerns over supply--demand balancing feasibility and costs. This article presents the first empirical analysis of speculative imbalance positioning (SIP) by NPTs in the GB electricity system through its effects on imbalance volumes, balancing costs, the imbalance price, and associated settlement costs over the period 2021--2024. Our research uses party-level imbalance data in each settlement period to run counterfactual simulations of the electricity balancing mechanism without the aggregate NPT imbalance. Our results indicate that SIP by NPTs reduced the total net balancing volume required by 3.1--3.9 TWh (16--20\%), reduced system imbalance standard deviation by 48--65 MWh (13--17\%), reduced balancing costs by £347--396m (5.9--6.7\%), and reduced supplier imbalance costs by £193--325m (65--75\%) over the analysis period. These results suggest that consumer bills would have been higher without SIP by NPTs, since the extra costs would likely have been passed on to consumers through levies, although the potential for effects in the wholesale (primarily late-stage intraday) markets complicates this picture.
\end{abstract}


\begin{highlights}
\item Non-physical traders' speculative imbalance positions reduce GB system imbalances
\item Total absolute imbalance volume was 16--20\% lower than without speculative positions
\item Balancing costs were £347--396m lower and supplier imbalance charges 65--75\% lower
\end{highlights}

%% Keywords
\begin{keyword}
Electricity market design \sep Electricity system balancing \sep Speculative trading \sep Electricity imbalance pricing



\end{keyword}

\end{frontmatter}



\section{Introduction}
\label{sec:Introduction}

Electricity system balancing in GB takes place primarily through the balancing mechanism (BM), in which NESO (the GB transmission system operator) simultaneously instructs both congestion and other transmission constraint management actions, as well as refining units' schedules to maintain aggregate supply and demand balance. Over the period 2021--2024, NESO spent £1.4 billion per year on average to manage supply and demand imbalances through the BM alone\footnote{Calculated from settlement stack data, based on flagged vs unflagged actions}. The average absolute imbalance over the same period in each half-hour window was 233 MWh, or around 4 TWh per year --- equivalent to 1.5\% of total energy delivered through the transmission network in 2024 \citep{neso2024review}. NESO has a licence condition to minimise consumer costs which highlights the need for efficient market mechanisms.

Trading parties in the GB electricity system are financially accountable for differences between their contracted and physical positions. In GB, along with other markets such as the Irish Single Electricity Market, the Netherlands, Denmark, and France, parties are allowed to take speculative positions on the imbalance price. This research assesses whether speculative imbalance positioning (SIP) by non-physical traders (NPTs) benefits the GB electricity system.

One key reason to permit SIP is to incentivise `helpful' imbalances \citep{Ofgem2015P305}, as explained in the following example. Suppose an NPT buys energy ahead of time, consumes no energy, but the system has a net upward balancing energy requirement anyway. This energy purchase likely induced an increase in generation, which reduces the net balancing energy requirement compared with the counterfactual of no purchase. The trade is profitable if the price at which the NPT's imbalance is settled (i.e. at which they sell their excess energy) exceeds the purchase cost. If the imbalance price is tied to the marginal cost of energy balancing, profitable imbalance arbitrage will also reduce the system imbalance \citep{BunnKermer}. However, whether this idealised scenario plays out in practice is an empirical matter. For example, if the settlement price (SP) does not reflect the marginal cost of buying balancing power, NPTs may be incentivised to take positions which exacerbate the system imbalance. Despite this trading strategy being common in GB \citep{Elexon}, no study has quantified its effects on the GB system. 

This research is timely for three main reasons. First, the UK government's review of electricity market arrangements (REMA) led to a decision to address the challenges through reformed national pricing (RNP). Among the latest set of proposals are measures that deliberately curb opportunistic imbalancing by small physical assets in real time (which we term passive balancing -- see Section~\ref{subsec:academicLiterature}), which they associate with adverse effects on grid stability and consumer costs \citep{nesoRnp}. These same measures would inhibit SIP, but (as the document acknowledges) there is no existing research on how these reforms would affect balancing volumes. Second, REMA states:

\begin{quote}
“Non-physical traders improve market liquidity and are an important component of a healthy market, therefore we want to encourage their participation” \citep{remaSecondConsultation}
\end{quote}
yet provides no evidence in the technical assessments supporting this position. Third, in 2020 ACER decided on single imbalance pricing as the standard methodology for the harmonisation of the imbalance settlement across member system operators (SOs) \citep{ACER}. Since this could create an incentive for opportunistic imbalancing across new markets, it is important to understand how single pricing has already played out in a market with significant structural similarities. This paper focuses on the GB system, due to the abundance of publicly available GB data on balancing. However, the analysis and methodology are also relevant to other energy systems, particularly those with liquid intraday markets, transparent imbalance price mechanisms, and a growing share of variable renewable energy sources in the generation mix, such as in Europe and Japan. 

This paper provides two main novel contributions. First, as far as the authors are aware this research represents the first significant published analysis on the role of NPTs in the GB system. Second, this research provides the first empirical quantification of the impacts of SIP on GB electricity balancing, ten years after the introduction of single imbalance pricing. 

In this research, we focus on the effects of SIP (by NPTs only) on balancing and imbalance settlement volumes and costs. Reducing these is one of the key aims of permitting SIP (see Section~\ref{sec:litReview}), and of significant public interest given the recent rise in balancing costs. Combining extensive publicly available data in new ways, we run counterfactual simulations of the GB balancing mechanism as the basis of our analysis. Over the period 2021--2024, our modelling suggests that SIP reduced the total volume of energy balancing actions required by 3.1--3.9 TWh (16--20\%), and reduced consumer bills through lower balancing costs to NESO (between £347--396m, or 5.9--6.7\%) and a reduction in imbalance charges to suppliers (between £193--£325m, or 65--75\%). While SIP also plays an important role in changing overall system emissions and short-term market price formation, estimating these effects either requires non-public information or ex-ante market modelling, which is outside the scope of this paper.

The remainder of the paper is structured as follows. Section~\ref{sec:litReview} covers GB trading arrangements and provides a literature review. Section~\ref{sec:methodologyAndData} covers the data sources and methodology of the analysis. Section~\ref{sec:results} reports the results on system imbalance volumes and financial effects. Finally, Section~\ref{sec:discussionAndPolicyImplications} discusses the implications of the results for ongoing policy discussions.

\section{Literature Review}
\label{sec:litReview}

This section provides an overview of GB electricity trading arrangements, and related academic literature on opportunistic imbalancing.

\subsection{GB Market Arrangements}
\label{subsec:marketArrangements}

\begin{figure}[htbp]
    \centering
    \includegraphics[width=\linewidth]{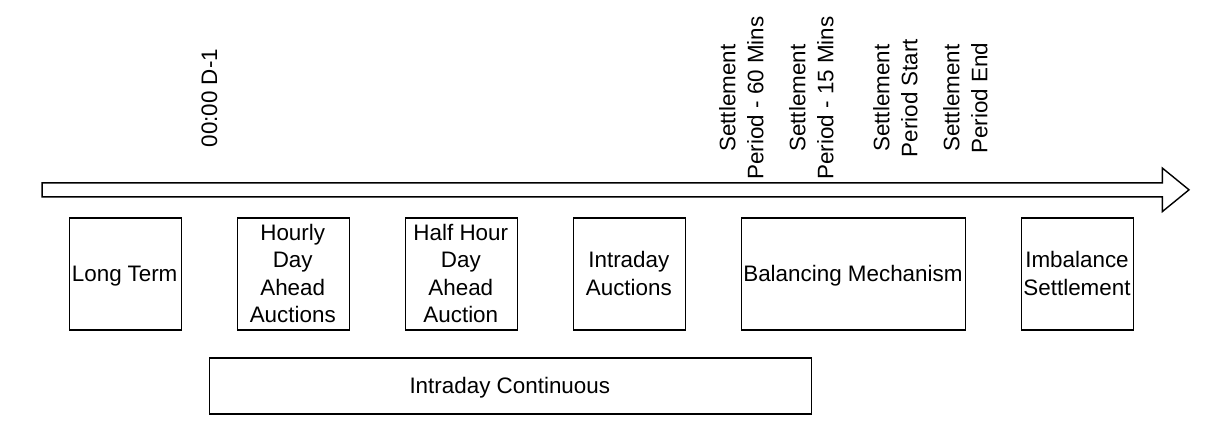}
    \caption{Timeline of trading arrangements in the GB electricity market. NESO manages supply and demand imbalances and system network constraints primarily within the BM, which starts 60 minutes before delivery. Parties may trade up to 15 minutes before delivery on the continuous intraday market, overlapping with the BM.}
    \label{fig:marketArrangements}
\end{figure}

Figure~\ref{fig:marketArrangements} provides an overview of domestic power trading arrangements in GB, via sequential ex-ante markets from years ahead until just before delivery. One hour before delivery (known as gate closure) large generators and demand units which participate in the BM submit their expected schedules and costs to deviate from their schedule. From gate closure until the end of a settlement period (the 30-minute blocks into which GB power days are divided), NESO balances the grid primarily through the BM. Some actions maintain exact supply/demand balance (known as energy balancing), while others manage transmission constraints such as line capacity limits (known as system balancing). In the 45-minute window between gate and intraday market closure, all trading parties can change their contracted positions through trading on the continuous intraday market, and non-BM units (such as smaller batteries and flexible plants) can change their output accordingly.

The difference between a party’s contracted and metered volume over the course of a settlement period is their imbalance volume. If they have oversupplied or under-consumed, their excess is sold back to the system at SP, and vice versa. The net volumetric sum of the actions taken by NESO in a settlement period is the NIV (net imbalance volume, see Equation~\ref{eq:nivDefinition}). SP is, generally, the price of the most expensive action taken by NESO in the direction of NIV, after netting off balancing actions taken in the opposite direction. For example, if NESO accepts 100 MWh of upward regulation and 20 MWh downward regulation in a settlement period, SP is set by the price of the 80\textsuperscript{th} MWh\footnote{There are various complications to this procedure, which are clearly detailed in \citep{ElexonImbalancePricingGuidance2025}.}. Changes to the total volume of balancing actions taken could affect SP, and hence cashflows to imbalanced parties.

In November 2015, the GB imbalance settlement switched from dual to single pricing. Among other reasons, Ofgem (the GB power market regulator) wanted to reward parties for taking `helpful' imbalances \citep{Ofgem2015P305}. The following example clarifies this rationale. Consider a part-loaded gas generator, who (correctly) identifies that the system will be short by 150 MWh of generation. Their marginal cost is $C$ £/MWh. Under a single imbalance price regime, SP is set by the marginal cost of upward regulation. If $C<SP$, the generator has an incentive of $(SP-C)$ £/MWh to generate more than their contracted quantity, since they receive SP for any excess power generated. If they do so, this reduces the system imbalance and saves the SO (and hence the consumer) at least $(SP-C)$ £/MWh compared to activating more costly upward balancing power. However, under a dual imbalance price regime, the price the generator receives for spilling into the system is often set by the \textit{downward} regulation marginal price. Since this generally falls below the marginal cost of gas generation, the part-loaded generator has little to no incentive to spill, and chooses not to do so. Instead, the SO has to activate (more costly) upward balancing power, increasing costs to the consumer. Effectively, this reform partly outsources energy balancing responsibility away from the BM, and encourages parties to anticipate and counteract the overall system imbalance.

\subsection{Academic Literature}
\label{subsec:academicLiterature}

Existing research falls into three main categories: imbalance price theory, modelling agent behaviour under different systems, and empirical analysis of the effects of changes to imbalance price regimes. This paper falls into the final category, filling a gap in the literature on exploring the effects of SIP in the GB system. In this review, it will be useful to distinguish between two different types of opportunistic imbalance positions:

\begin{description}
  \item[Passive Balancing:] Asset-backed traders voluntarily change their output from their contracted position to access a more favourable price at imbalance than available in ex-ante markets. For example, a small battery changes its output in response to the latest publications of balancing actions taken by the SO.
  \item[Speculative Imbalance Positioning:] NPTs speculating on price spreads between short-term market prices and the SP. For example, in anticipation of the SP exceeding the current intraday price, an NPT might purchase power at intraday, without any intention to consume the power purchased.
\end{description}

Both strategies aim to profit from predicting SP. From a system perspective, these strategies can coincide in terms of their physical effects. For example, an NPT may contract with a small asset post-gate closure, effectively transferring the SP risk from the asset to the NPT, but with the same net effect as the small asset spilling into the system. However, some SIP occurs pre-gate closure, which can provide visibility to the SO if the counterparty submits their generation schedule in advance.

Combining single imbalance pricing with allowing opportunistic imbalancing has theoretical support. Suppose the forward markets settle to some equilibrium price, but forecast errors mean the system is short. Here, the SO is buying marginal power. In GB, where SP is generally reflective of the marginal costs of buying power \citep{Chen}, a long position from the forward markets displaces a higher priced balancing action, thus reducing balancing costs and SP (albeit at a loss of producer surplus at the margin) \citep{BunnKermer, bunnInekweMacGeehan}. Other research proposes alternative imbalance price and settlement schemes, such as in \citep{ghiassiFarrokhfalEtAl, lago}. However, these alternatives are beyond the scope of this research.

There exists significant simulated evidence that opportunistic imbalancing can be a profitable strategy in single price regimes. For example, a risk-averse opportunistic imbalance strategy backtested on the Belgian market would have been profitable \citep{bruneelEtAl}, while a passive balancing strategy in the same market yields similar outcomes for a BESS operator \citep{madahiEtAl}. Density forecasts are incorporated in a passive balancing strategy for a gas-fired unit in the Austrian market in \citep{bunnGianfredaKermer}, which offers improvements over techniques based on traditional expected value forecasts. Even simple trading strategies based on predicting only the sign of the imbalance, such as those in \citep{lisiEdoli, PozzettiCartlidge2020}, can generate profits through opportunistic imbalancing in the Italian and GB markets respectively. However, a risk-constrained strategy in GB outperformed a similar simple benchmark \citep{browell}.

While incentives for opportunistic imbalancing under dual pricing are generally lower than under single pricing, one simulation in Denmark suggests that even under dual pricing, an asset-backed player (a wind farm in this case) is incentivised to take a passive balancing position \citep{chavesAvila}. However, different price-setting rules in the Netherlands (historically) suggest that arbitrage profits were negative on average under dual pricing \citep{boogertDupont}. 

Only a few papers move beyond the benefits for the individual to consider the effects on overall system balancing in a European-style market (namely with day-ahead, continuous intraday, balancing, and imbalance settlement stages), which is one of the key contributions of our research. One study on a generic European-style market finds that overall imbalance costs are lowest under single price incentives \citep{vanDerVeen}. In the Austrian market, NPTs are simulated to take system-supportive and profitable imbalance positions, provided information relevant to forecasting is published sufficiently promptly \citep{BunnKermer}. A simulation on the Japanese market shows that despite the dual pricing disincentive against opportunistic imbalancing, profitable and system-supportive arbitrage opportunities remain \citep{Matsumoto}. However, a trading algorithm which takes speculative intraday positions based on expected imbalance prices, backtested on German data between 2017 and 2019, finds that a profitable trading strategy would not have been system supportive due to biased imbalance price incentives \citep{Koch}. A more efficient intraday price constraint would make this strategy more system supportive by tying the imbalance price to the true marginal cost of procuring balancing energy.

In practice, the following studies provide strong evidence that balancing parties respond to imbalance price incentives, regardless of whether opportunistic imbalancing is permitted. Following a switch from a single to a hybrid single/dual imbalance pricing regime in Italy, imbalances from demand-side balance responsible parties fell drastically \citep{cloFumagalli}. An analysis of the German grid suggests that taking a systematically short position through passive balancing would have been highly profitable. However, in 2016--17 there was no systematic arbitrage opportunity left, as market players systematically repositioned their traded volumes in response to asymmetric penalties for being short when the system is short vs. long when the system is long \citep{KochHirth}. Later research finds a significant association between the demand for imbalance energy (i.e. aggregate imbalance) and the imbalance price in Germany. An increase in the imbalance price is associated with a \textit{decrease} in demand for imbalance energy \citep{eicke}, despite the prohibition of strategic imbalances in this market.

Another study investigates whether parties take opportunistic imbalance positions, and what the effect on system balancing is. The analysis suggests traded volumes react to the publication of the latest system balance. These changes to traded positions are found to reduce system balancing volumes as well as the 95th percentile of system imbalances \citep{KochMaskos}. A different approach reaches a similar conclusion, finding significant evidence of statistical arbitrage through the correlations between DA and imbalance price spreads. There is no evidence that this practice threatened system stability compared with other systems \citep{moeller}. 

Other research casts doubt on the benefits of passive balancing. Over the period 2009--2011 in Germany, imbalance prices were relatively predictable, but not strongly correlated with spot (day-ahead and intraday) prices due to the market design \citep{JustWeber}. The mechanism allowed for dominant strategies by generators which threatened system stability. In practice, some players exploited this system which destabilised the grid by exacerbating imbalances, and cost consumers through raised spot prices and higher reserve capacity requirements.

Finally, Elexon (the balancing and settlement administrator in GB) investigate increases in imbalance volumes by individual parties following the switch to single imbalance pricing \citep{Elexon}. While imbalances by generators and suppliers remained broadly flat over the period 2014--2019, imbalances by NPTs grew significantly. They find that most imbalances by NPTs are in the same direction as the market imbalance, i.e. when the system is long, NPTs are also long. However, the growth in NPT imbalances has not led to an increase in balancing energy volumes, nor the most extreme NIVs. We return to this counter-intuitive finding in Section~\ref{subsec:imbalanceVolumes}, arguing that this is the wrong metric to understand the effect of SIP on system imbalances. Further investigation in \citep{elexonNivChasingFollowUp} finds evidence that NPTs likely take speculative imbalances in the GB market, but does not estimate the system effect of this strategy.

Beyond the European/Japanese context, there exists an extensive literature on the role of speculators in arbitraging day-ahead and real-time price spreads in US markets. Many researchers find that speculation has benefited US power systems by curbing generator market power, promoting price convergence, and reducing system operating costs and carbon emissions (see, e.g., \citep{jhaWolak2019, jhaWolak2023, mercadal, saravia}). However, others find that speculation can increase system operational complexity \citep{parsonsEtAl}, or increase spreads to gain higher revenues from power derivatives \citep{birgeEtAl}. We do not discuss this literature in detail, because US markets differ significantly from European-style markets (and hence from our setting) for two key reasons. First, the clearing mechanism in the US is based on central dispatch and nodal pricing, whereas most European markets rely on self-dispatch (to some extent) and zonal or national pricing. Second, US markets do not have a continuous intraday market, meaning speculation occurs at day-ahead. The ability to speculate closer to delivery with better information and the salient differences in market clearing mean that our institutional setting is sufficiently distant from the US that conclusions do not necessarily carry over.

In light of the mixed findings from other markets on the benefits of opportunistic imbalancing, this subject merits closer analysis. This research aims to fill the gap in the literature in the GB context, and extend the analysis of system balancing implications using counterfactual simulations incorporating real data to generate estimated cost effects. 

\section{Data, Assumptions, and Methodology}
\label{sec:methodologyAndData}

The overall aim of this research is to estimate the effects of SIP on BM volumes and costs by setting up a counterfactual scenario without SIP, and to use this analysis to infer the effects on balancing costs and imbalance settlement. This section discusses the data required, key assumptions in the analysis, and finally highlights our novel methodology.

\subsection{Data}
\label{subsec:data}

Elexon categorises balancing parties under different but not mutually exclusive labels. In this analysis, we define NPTs, generators, suppliers, and mixed-role parties as in Table~\ref{tab:party_categorisation}. These categories allow us to understand how SIP affects the imbalance cashflows of different system participants.

\begin{table}[htbp]
\centering

\begin{tabular}{l C{3cm} C{3cm} C{3cm}}
\toprule
\textbf{Category} & \textbf{TN (NPT)} & \textbf{TS (Supplier)} & \textbf{TG (Generator)} \\
\midrule
NPT       & \cmark & \xmark & \xmark \\
Generator & \xmark & \xmark & \cmark \\
Supplier  & \xmark & \cmark & \xmark \\
Mixed Role & \multicolumn{3}{>{\centering\arraybackslash}m{9cm}}{More than one of TN, TS, and TG} \\
\bottomrule
\end{tabular}
\caption{The definitions of the four main types of trading parties in our analysis, according to their registration status with Elexon.}
\label{tab:party_categorisation}
\end{table}

These definitions exclude opportunistic imbalances by asset-backed firms who trade beyond their physical volumes (essentially non-physical trading). However, it is difficult to distinguish the effects of these actions from passive balancing or unintentional mismatches between traded position and metered volume (e.g. wind forecast errors or outages). This is a significant caveat of our analysis --- the total effect of opportunistic imbalances may be larger than quantified in this research. However, we aim to investigate purely SIP by NPTs, whose trading model is primarily arbitrage rather than managing a physical portfolio, and as such are likely to represent the majority of pure speculative imbalance volumes.

We analyse the following two counterfactuals:

\begin{description}
    \item[Any Metered Volume (AMV):] All NPT imbalances are considered deliberate speculative volumes regardless of the party's metered volume. An NPT's physical volume may be non-zero if, for example, they trade over interconnectors, or optimise an asset on behalf of another party. This represents an upper bound on total SIP volumes, discussed below.
    \item[Zero Metered Volume (ZMV):] Only NPTs for whom their period metered volume is 0 are included. In any given period, these NPTs are a subset of all NPTs included in the AMV scenario. As such, this counterfactual represents a lower bound on total SIP volumes.
\end{description}

The study covers 2021--2024. There are 48 settlement periods in a day (except clock change days, in which there are 46 in short days and 50 in long days), which gives a maximum of 70,128 data points. We remove settlement periods in which data was partially or fully missing from the analysis, losing 1,305 settlement periods (roughly 27 days’ worth of data). Most of these points are due to market index price data missing, predominantly in 2021, and physical BM data missing in September 2022, leaving 68,823 data points. The sources for all datasets used in this analysis can be found in the replication guide in the online GitHub repository (\url{https://github.com/PSALOxford/niv-chasing-bm-analysis/tree/main}).

\subsection{Assumptions}
\label{subsec:assumptions}

Underlying this research is the assumption that SIP does not affect the imbalances of other parties. In other words, this implies that a supplier would have found another counterparty with whom to trade in the counterfactual, such that their overall imbalance (difference between traded and physical positions) is unchanged, or in the case of a generator, there is simply less generation on the system. Anecdotally, generators aim to match their expected output with their traded positions, unless they themselves are passive balancing (such as smaller batteries). This assumption is one of the main limitations of our work for two key reasons. First, without SIP, we would expect lower liquidity post-gate closure. This is a crucial period of the intraday market for renewable generators looking to cover an imbalance implied by forecast updates. However, this conservative assumption is in keeping with underestimating the true benefits of SIP throughout this paper. Second, it is unlikely that these assets would take the same passive imbalance positions in the counterfactual, since the real time state of the system would be different. However, it is impossible to know what these passive balancing volumes are, so we do not attempt to account for this. On the supplier side, it is unlikely they would be able to purchase the same volumes in the different scenarios, since one of the key functions of NPTs is to provide liquidity. However, this assumption should affect only the supplier imbalance benefit part of our analysis, and not the overall picture on balancing volumes and costs.

\begin{figure}[htbp]
    \centering
    \includegraphics[width=\linewidth]{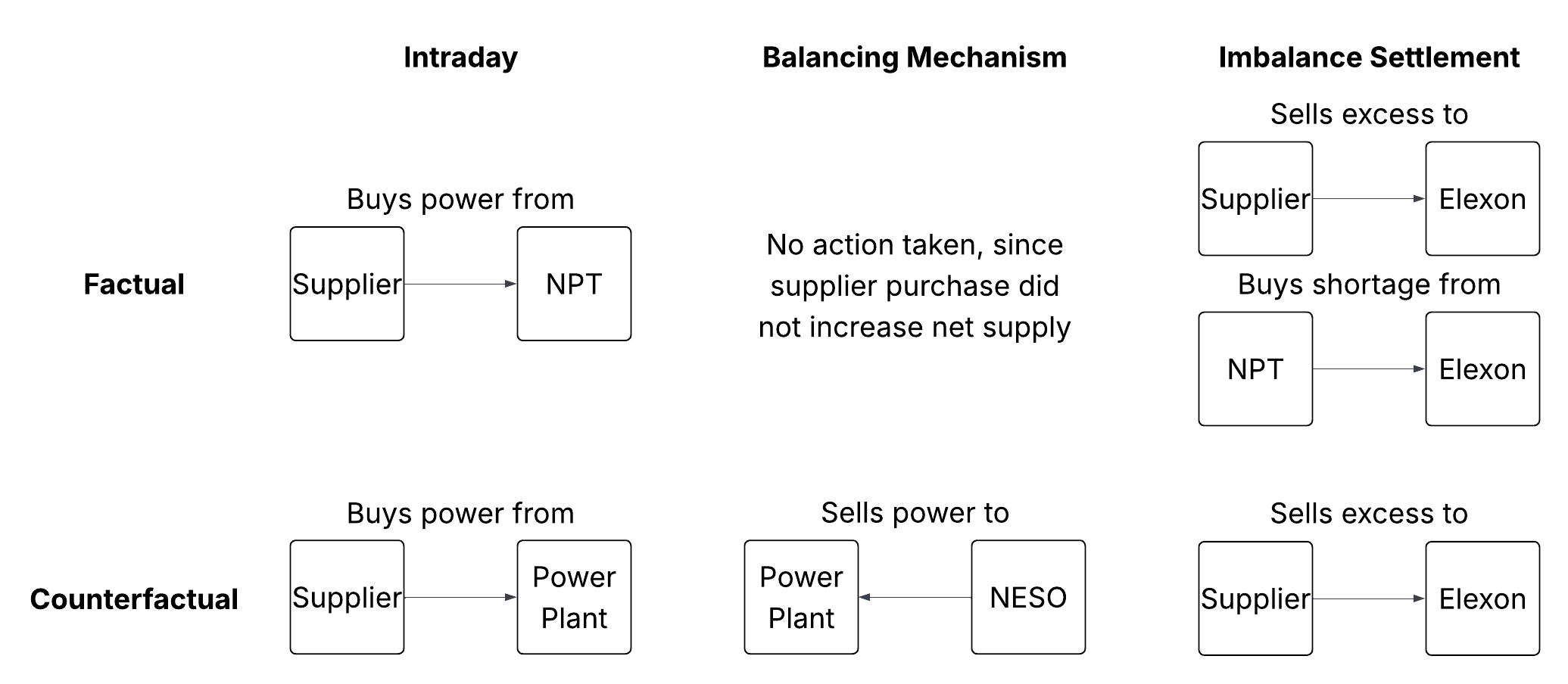}
    \caption{Stages of SIP for a scenario in which the system ends up with excess power due to suppliers over-forecasting demand, demonstrating the differences in generation, system balancing, and imbalance settlement between the factual (with SIP) and the counterfactual (without).}
    \label{fig:nivChasingStages}
\end{figure}

Figure~\ref{fig:nivChasingStages} is an example of this assumption. Suppose a supplier over-forecasts their demand, and purchases power at intraday. In the factual scenario (where NPT is allowed), an NPT might sell this contract, but without backing the contract with physical output. In the counterfactual scenario (where NPT is disallowed), the supplier might purchase the same amount of energy from an asset-backed generator, thus increasing scheduled physical system supply. Suppose the system has excess power. In the counterfactual case, the SO would have to accept another bid to decrease supply, compared to the factual case in which the NPT simply buys their shortage from the settlement administrator. 

For our analysis, we make three further key assumptions.

First, we assume that the stack of available actions in the BM is unchanged between the factual and counterfactual scenarios. This does not provide a perfect simulation of the counterfactual: clearly, if NPTs cannot take speculative positions, and suppliers instead purchase the same energy directly from generators, then the same generator will not submit the same prices in the BM as they would in the factual case. However, we do not expect that this assumption will alter our conclusions significantly. Anecdotally, most speculative trades occur after gate closure, i.e. when BM units have already submitted their physical positions, bids, and offers to the SO. Hence most physical effects of SIP occur due to interactions with smaller plants anyway, which do not interact with the BM significantly. 

Second, our simulation assumes the SO takes balancing actions through the BM (rather than ancillary services) in merit order. This means the results in this paper \textit{underestimate} the effects of SIP on balancing costs and hence the settlement price, since the SO does not always accept balancing actions in merit order, often because doing so would violate an existing transmission constraint.

Third, we assume all balancing actions are flagged correctly (we discuss the importance of `flagging' further in Section~\ref{subsubsec:methods-BmSimulation}). The SO flags actions which address a system management constraint (e.g. thermal line capacity), rather than energy balancing, since the BM handles both balancing responsibilities. In our simulation, all SO-flagged actions are assumed to be taken in the counterfactual, since we assume NESO would have taken the same system management actions in the counterfactual (i.e. the level of SO-flagged actions required is unchanged). The accuracy of this flagging procedure was $\geq$99.5\% in FY2024--25 \citep{nesoFlaggingReport}, so we anticipate this assumption will have a very limited impact on the results. 

\subsection{Methods}
\label{subsec:methods}

In this section, we discuss the methods used to calculate NIV, BM costs, SP, and imbalance settlement costs in the counterfactual scenario without SIP. A detailed guide on workflow replication can be found in the accompanying GitHub repository. Table~\ref{tab:methodologySymbols} provides a key for the symbols used in this section.

\begin{table}[htbp]

  \centering
  \footnotesize
  \begin{tabularx}{\textwidth}{c X} 
    \toprule
    \textbf{Symbol} & \textbf{Description} \\
    \midrule
    $N$ & GB trading parties classified as NPTs according to Table~\ref{tab:party_categorisation}, a subset of all GB trading parties\\
    $S$ & GB trading parties classified as suppliers according to Table~\ref{tab:party_categorisation}\\
    $G$ & GB trading parties classified as generators according to Table~\ref{tab:party_categorisation}\\
    $M$ & GB trading parties classified as mixed role according to Table~\ref{tab:party_categorisation}\\
    $I_p$ & Imbalance (MWh) of party $p$\\
    $q_i$ & Volume (MWh) of accepted balancing action $i$ in a given settlement period (positive is offer; negative is bid)\\
    $p_i$ & Price (£/MWh) associated with accepted balancing action $i$\\
    $C_p$ & Imbalance cashflow (£) for party $p$\\
    $BM$ & Total BM costs (£) in a given settlement period\\
    \bottomrule
  \end{tabularx}
  \caption{Definitions of symbols used in the methodology}
  \label{tab:methodologySymbols}
  
\end{table}

\subsubsection{Counterfactual Imbalance Volumes}
\label{subsubsec:recalculatingVolumes}

We investigate the effects on imbalance volumes by calculating NIV in each settlement period without NPTs. First, we calculate an aggregate imbalance by NPTs from reports available upon request from Elexon which provide the traded volume, metered volume, and energy imbalance volume for each trading party in each settlement period.

\begin{equation}
    Imb = \sum_{n\in N}I_n
    \label{eq:imbalanceDefinition}
\end{equation}

$NIV$ is the net volume of balancing actions taken:

\begin{equation}
    NIV = \sum_iq_i
    \label{eq:nivDefinition}
\end{equation}

However, a positive party imbalance implies the party is oversupplied. Hence, the counterfactual $NIV$ ($NIV^\prime$) is calculated as:

\begin{equation}
    NIV^\prime=NIV+Imb
\label{eq:counterfactualNivDefinition}
\end{equation}

\subsubsection{Balancing Mechanism Simulation}
\label{subsubsec:methods-BmSimulation}

\begin{figure}[!t]
    \centering
    \includegraphics[width=\linewidth]{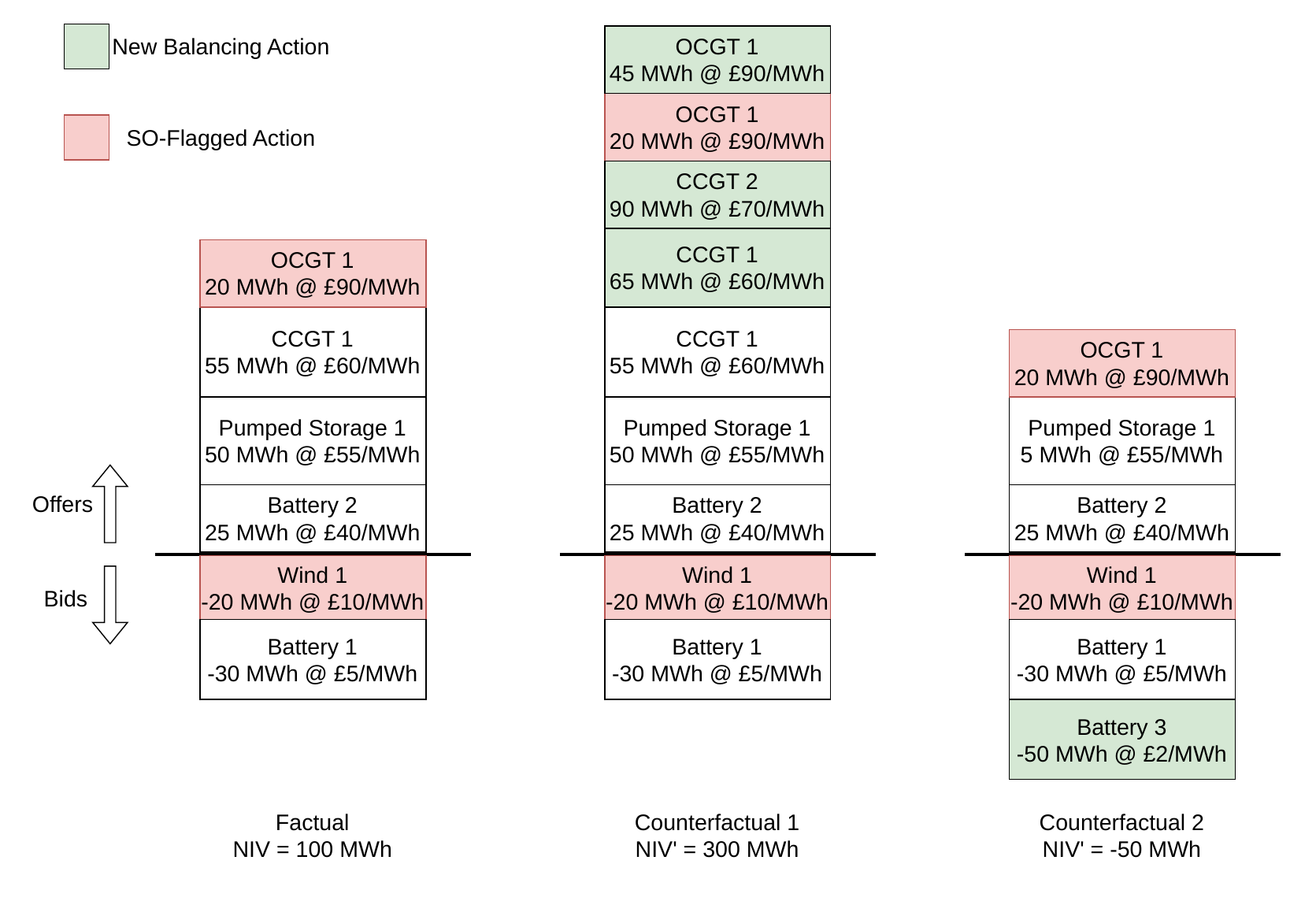}
    \caption{Schematic of merit order simulation for two counterfactual imbalances: one in which the SO requires extra upward regulation, and another in which accepted volume must be removed and further bids accepted.}
    \label{fig:BmSimulation}
\end{figure}

In our counterfactual BM simulation, we aim to model the balancing actions the SO would have taken if the overall system imbalance were $NIV^\prime$, i.e. without SIP. Figure~\ref{fig:BmSimulation} shows two examples of the simulation we run. In both cases, $NIV = 100$ MWh. In counterfactuals 1 and 2, $Imb = 200$ and $-150$ MWh respectively, so $NIV^\prime = 300$ and $-50$ MWh.

In counterfactual 1, we accept $200$ MWh of upward regulation. We assume the SO accepts offers starting with the marginal unflagged offer, in this case from CCGT 1. If this plant has headroom, the SO accepts the remaining volume at this price, before moving to the next cheapest offer (CCGT 2) to fulfil the next portion of volume requirement. This merit-order process repeats until the total accepted volume meets the $NIV^\prime$ requirement. We do not remove accepted bids in this instance, on the assumption that these bids were required to manage short-term fluctuations in supply and demand.

In counterfactual 2, we reduce the imbalance volume by $150$ MWh. To do so, the most expensive unflagged offers are removed, until the volume of offers matches the volume of bids. Then, the SO accepts the highest priced bids, starting with the current marginal downward regulation plant (in this case Battery 1). In this example, we suppose Battery 1 has no footroom, and as such the SO accepts a bid from the next highest priced unit, Battery 3. As before, SO-flagged actions are unchanged between the factual and counterfactual. Pseudocode for our simulation engine is available in~\ref{app:pseudocode}.

\subsubsection{Financial Effects}
\label{subsubsec:financialEffects}

The counterfactual BM costs to the SO are calculated directly from the new settlement stack as:

\begin{equation}
    BM = \sum_iq_ip_i
\end{equation}

From the counterfactual stack of balancing actions (e.g. Counterfactuals 1 and 2 in Figure~\ref{fig:BmSimulation}), we use Elexon's procedures to calculate $SP^{\prime}$ \citep{Elexon2025SectionT, ElexonImbalancePricingGuidance2025}. Elexon aims to calculate a price which reflects the cost of managing supply-demand imbalances over the 30-minute settlement period. As such, not all balancing actions contribute to $SP$ --- for example, actions which manage transmission constraints, or short duration/very small volume actions do not set the imbalance price. The full calculation is not detailed here since it is not an original part of our methodology, but the algorithm is implemented in the supporting GitHub repository and replicated in pseudocode in~\ref{app:pseudocode}. The algorithm written to calculate $SP$ from balancing actions was backtested on 2024 data to ensure accuracy.

Using $SP^{(\prime)}$ we calculate the imbalance settlement charges/payments to each party, which are just signed net cashflows through the imbalance settlement mechanism:

\begin{equation}
    C_p^{(\prime)} = I_p\cdot SP^{(\prime)}
\label{eq:imbalanceChargeDefinition}
\end{equation}
where positive is a payment to the party. In calculating aggregate cashflows for each of the types of trading parties in Table~\ref{tab:party_categorisation} (supplier, generator, or mixed role), we sum the cashflows for each trading party in that type. For example, supplier cashflows are defined as:

\begin{equation}
    C_S=\sum_{p\in{S}}C_p
    \label{eq:imbalanceCashflowByCategoryFormula}
\end{equation}

\section{Results}
\label{sec:results}

Our study covers each half-hour period from 2021 to 2024, excluding periods in which relevant data was missing. A summary of the key results can be found in Table~\ref{tab:resultsSummary}, discussed in greater detail in the subsections to follow.

\begin{table}[htbp]
  \centering
  \footnotesize
  \begin{threeparttable}
  \begin{tabularx}{\textwidth}{Y Z Z Z Z Z} 
    \toprule
    \textbf{Metric (unit)} & \textbf{Factual} & \textbf{AMV} & \textbf{ZMV} & \textbf{\makecell{AMV -\\Factual}} & \textbf{\makecell{ZMV -\\Factual}} \\
    \midrule
    Average NIV (MWh) & 1.9 & -26.0 & -16.0 & -27.9 & -17.9\\
    Average absolute NIV (MWh) & 233 & 278 & 291 & 44.3 & 57.3\\
    Total absolute NIV (TWh) & 16.1 & 19.1 & 20.0 & 3.1 & 3.9\\
    NIV standard deviation (MWh) & 312 & 361 & 378 & 48.1 & 65.3\\
    Max NIV (MWh) & 1737 & 1792 & 2281 & 54.5 & 544\\
    Min NIV (MWh) & -1869 & -1925 & -1936 & -55.5 & -66.7\\
    \% |imbalances| $>1$ GWh & 0.71 & 1.21 & 1.44 & 0.50 & 0.73\\
    Total BM costs (£bn) & 9.00 & 9.40 & 9.35 & 0.40 & 0.35\\
    Average system price (£/MWh) & 119.8 & 121.3 & 119.2 & 1.47 & -0.62\\
    Total SO imbalance settlement cashflow (£m) & 853 & 959 & 1032 & 106 & 179\\
    $S$ imbalance cashflow (£m) & -106 & -431 & -299 & -325 & -193 \\
    $G$ imbalance cashflow (£m) & -221 & -280 & -263 & -60 & -42\\
    $M$ imbalance cashflow (£m) & 93 & -69 & -0.9 & -163 & -94\\
    \bottomrule
  \end{tabularx}
  \end{threeparttable}
  \caption{Comparison of factual and modelled system imbalance metrics. A positive NIV implies the system was short. All financial results are in nominal terms.}
  \label{tab:resultsSummary}
\end{table}

\subsection{Imbalance Volumes}
\label{subsec:imbalanceVolumes}

\begin{figure}
    \centering
    \includegraphics[width=\linewidth]{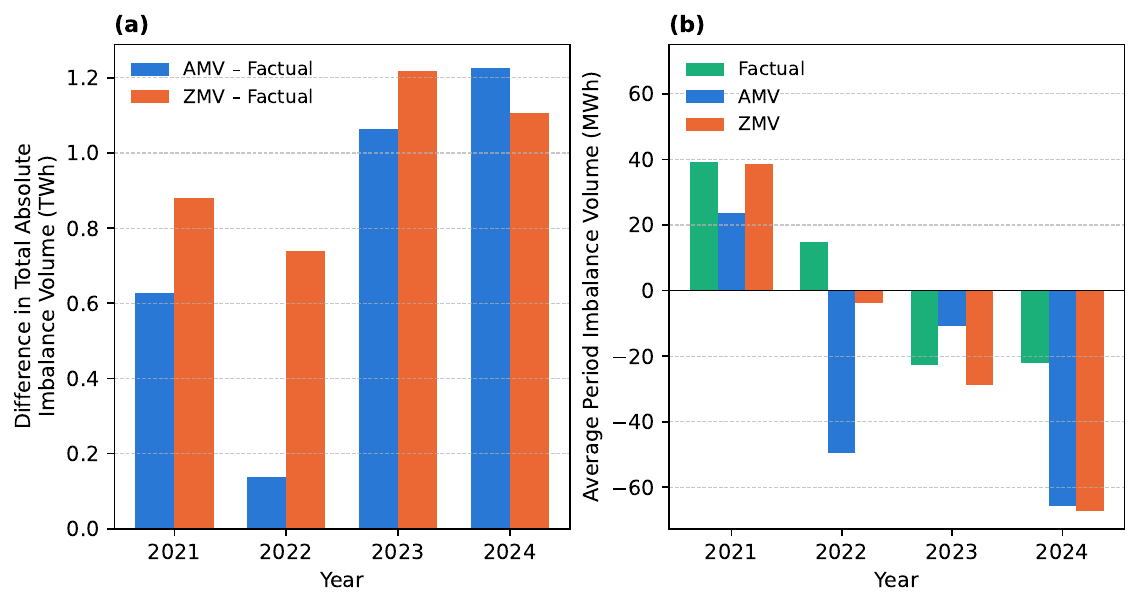}
    \caption{(a) Bar chart showing the difference in the total absolute volume of balancing actions required by year between the counterfactual without SIP (AMV/ZMV cases) vs the factual, with NPTs. (b) Bar chart showing the effect of SIP on the average period imbalance by year, comparing the factual and counterfactual scenarios.}
    \label{fig:absImbalanceDiffs}
\end{figure}

\begin{figure}[htbp]
    \centering
    \includegraphics[width=\linewidth]{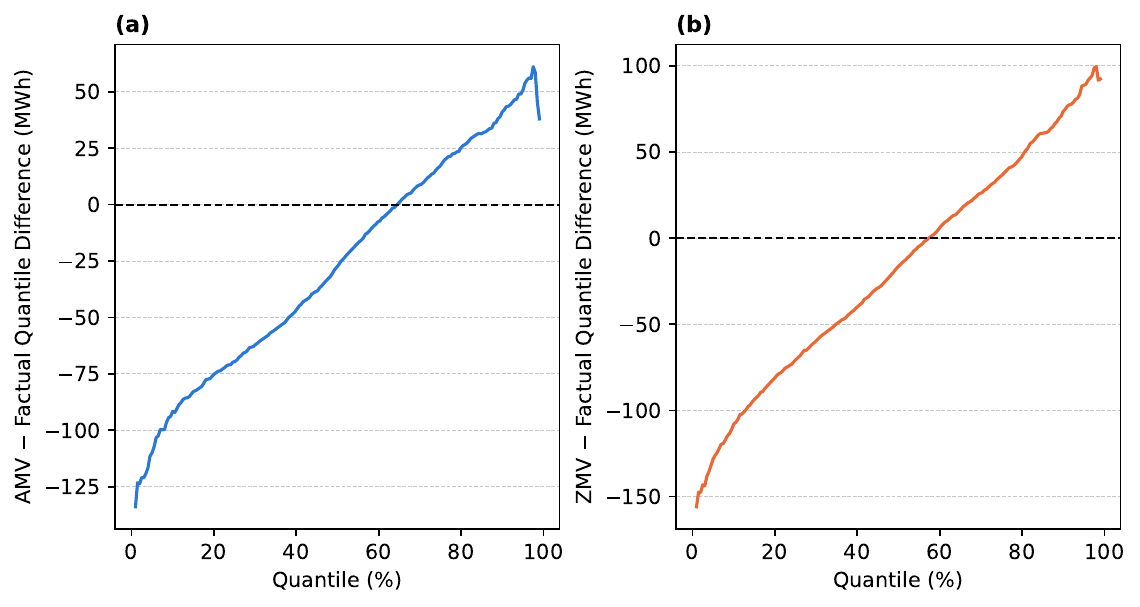}
    \caption{Quantile difference between counterfactual (without NPTs) and factual (with NPTs) net imbalance volumes 2021--2024. Only the 1\textsuperscript{st}--99\textsuperscript{th} percentiles are displayed for clarity of representation.}
    \label{fig:quantileDiff}
\end{figure}

We run the two scenarios AMV (any metered volume) and ZMV (zero metered volume) explained in Section~\ref{subsec:data} in our analysis. These aim to bound our estimate of the true effects of SIP through the net effect of all NPTs (AMV) and pure SIP (ZMV). Our results indicate that SIP reduced the total volume and volatility of required balancing actions, as shown in Figures~\ref{fig:absImbalanceDiffs} and \ref{fig:quantileDiff} which we discuss now. 

Figure~\ref{fig:absImbalanceDiffs}(a) demonstrates the effect of shifting otherwise required balancing volumes into the wholesale markets on the total absolute volume of balancing actions required over 2021--2024. Notably, in each year, the total volume of balancing actions NESO would have had to take would have been higher without SIP in both the AMV and ZMV cases. Figure~\ref{fig:absImbalanceDiffs}(b) shows the effect of SIP on the average imbalance volume in each settlement period. This highlights that SIP can significantly alter both the sign and magnitude of the average raw imbalance. The financial implications of these results are explored in Section~\ref{subsec:financialEffectsResults}. 

Figure~\ref{fig:quantileDiff} demonstrates the effect of SIP on the volatility and tails of the imbalance distribution  through a quantile difference plot for the AMV (Figure~\ref{fig:quantileDiff}(a)) and ZMV (Figure~\ref{fig:quantileDiff}(b)) scenarios. These results indicate that the tails of the imbalance distribution (below/above the 10\textsuperscript{th}/90\textsuperscript{th} percentiles) are more extreme in the counterfactuals. This suggests SIP reduces the frequency of extreme events, which is important for balancing reserve sizing. Both plots reveal a clear positive slope, which (in our case) indicates a more volatile imbalance volume in the counterfactual.

Beyond the summaries above, we present a high-level summary of the effect of SIP on balancing volumes in Table~\ref{tab:imbalanceVolumeSummary}. In this table, we present the proportion of periods in which the aggregate NPT imbalance interacted with the overall system imbalance under our different scenarios. The average NPT position indicates that in both scenarios, NPTs are net sellers in the intraday market and hence take an average short position through to imbalance. The average \textit{absolute} position is of a comparable magnitude to the average absolute factual NIV (see Table~\ref{tab:resultsSummary}). A naive comparison of NPT imbalances would find that these do not oppose the outturn NIV in the AMV case on average, and only do so 56.5\% of the time in the ZMV case. This could indicate that SIP fails to anticipate NIV correctly, and as such does not contribute to system balancing. However, this would be an inappropriate reading of the results. In both cases, NPTs correctly anticipate the sign of the system imbalance \textit{without their intervention} around 70\% of the time. The difference between these two results lies in multiple NPTs attempting to defeat the same imbalance. Most importantly, even with competition among NPTs flipping the sign of the imbalance, the absolute counterfactual NIV would have been larger without NPTs more often than not in both cases. This explains the observed reduction in absolute imbalance volumes observed in Table~\ref{tab:resultsSummary}.

\begin{table}[htbp]
  \centering
  \footnotesize
  \begin{threeparttable}
  \begin{tabularx}{\textwidth}{X c c} 
    \toprule
    \textbf{Metric (unit)} & \textbf{AMV} & \textbf{ZMV} \\
    \midrule
    Average NPT position (MWh) & -27.9 & -17.9\\
    Average absolute net NPT position (MWh) & 198 & 123\\
    NPT position standard deviation (MWh) & 254 & 163\\
    Max aggregate imbalance position (MWh) & 1477 & 1946\\
    NPT imbalance opposed to \textit{factual} system imbalance (\%) & 43.1 & 56.5\\
    NPT imbalance opposed to \textit{counterfactual} system imbalance (\%) & 68.0 & 71.0\\
    Absolute counterfactual imbalance > absolute factual imbalance (\%) & 57.4 & 65.0\\
    Sign of counterfactual imbalance $\neq$ sign of factual imbalance (\%) & 25.0 & 14.4\\
    \bottomrule
  \end{tabularx}
  
  \end{threeparttable}
  {\caption{Summary of the interaction between NPT imbalance volumes and factual/counterfactual imbalance volumes.}
  \label{tab:imbalanceVolumeSummary}
  }
\end{table}

% \begin{table}[htbp]
%   \centering
%   \footnotesize
%   \begin{threeparttable}
%   \begin{tabularx}{\textwidth}{X c c} 
%     \toprule
%     \textbf{Metric (unit)} & \textbf{AMV} & \textbf{ZMV} \\
%     \midrule
%     Average NPT position (MWh) & -27.9 & -17.9\\
%     Average absolute net NPT position (MWh) & 198 & 123\\
%     NPT position standard deviation (MWh) & 254 & 163\\
%     Max aggregate imbalance position (MWh) & 1477 & 1946\\
%     NPT imbalance in opposite direction to \textit{factual} system imbalance (\%) & 43.1 & 56.5\\
%     NPT imbalance in opposite direction to \textit{counterfactual} system imbalance (\%) & 68.0 & 71.0\\
%     Absolute counterfactual imbalance > absolute factual imbalance (\%) & 57.4 & 65.0\\
%     Sign of counterfactual imbalance opposite to sign of factual imbalance (\%) & 25.0 & 14.4\\
%     \bottomrule
%   \end{tabularx}
%   \end{threeparttable}
%   \caption{Summary of the interaction between NPT imbalance volumes and factual/counterfactual imbalance volumes.} 
%   \label{tab:imbalanceVolumeSummary}
% \end{table}

\subsection{Financial Effects}
\label{subsec:financialEffectsResults}

By reducing the balancing volumes required, SIP affects the costs NESO incurs through the BM, and imbalance settlement costs for other parties through changes to SP. While there will also be financial consequences in the ex-ante markets, we do not provide estimates for this within this research. Doing so would shift the focus from using publicly available data to support precise calculations towards estimating impacts on day-ahead and intraday markets, which would be useful future research. We find that SIP brings significant overall financial benefits in the BM and at imbalance settlement.

\begin{figure}
    \centering
    \includegraphics[width=\linewidth]{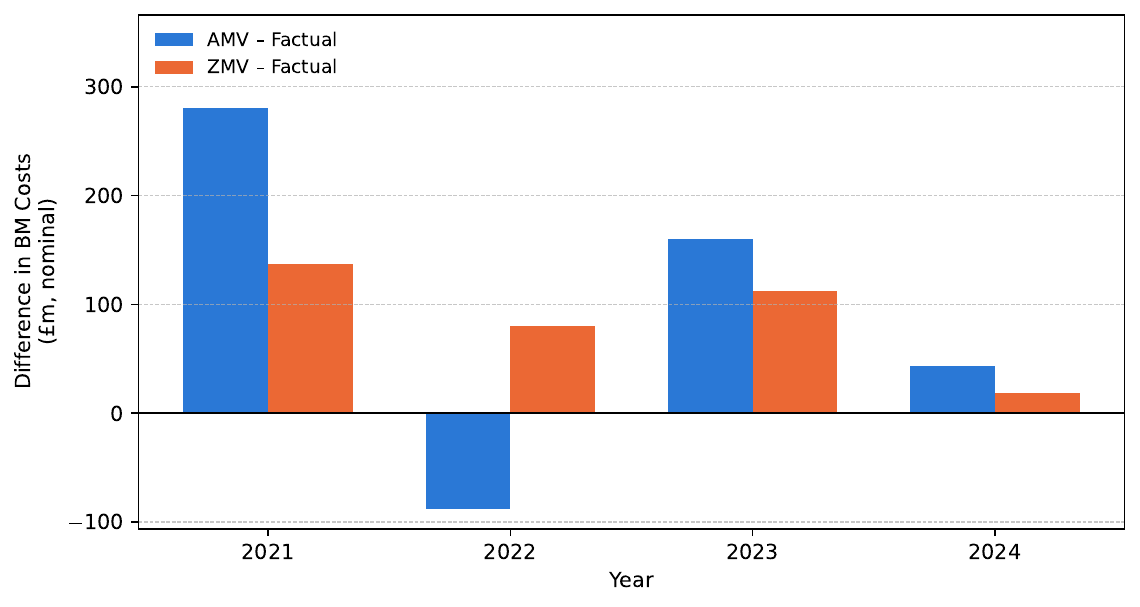}
    \caption{Bar chart plotting the differences in BM costs between AMV and ZMV scenarios and the factual (with SIP) by year.}
    \label{fig:bmCostsDifference}
\end{figure}

Figure~\ref{fig:bmCostsDifference} plots the simulated difference in BM costs across the scenarios. Overall, SIP reduces balancing costs by between £396m (AMV) and £347m (ZMV), a significant (5.9--6.7\%) reduction in the overall spend on SO-unflagged actions in the BM over the period 2021--2024 (£5.54bn). However, the plot reveals significant differences by year. Most notably, the AMV case appears to represent a net cost to the SO in 2022, while the benefits in 2024 are significantly smaller than other years. To understand this heterogeneity, we return to Figure~\ref{fig:absImbalanceDiffs}(b). In 2022, the average imbalance in the AMV case was -49.5 MWh, compared with 15.0 MWh in the factual case. Although SIP reduced the total imbalance volume in 2022-AMV (Figure~\ref{fig:absImbalanceDiffs}(a)), the average imbalance flipped from long (AMV) to short (factual). As such, the SO went from receiving payments to reduce generation, to paying to increase generation. 2024 is a similar story: average imbalance in AMV and ZMV cases was significantly longer, meaning the SO received lower revenues for reducing output. This motivates the need to look at imbalance costs as well as balancing costs to understand the true effect of SIP: although the SO (and hence consumers) might appear to benefit from lower balancing costs, oversupplied retailers would have received lower imbalance settlement revenue.

Suppliers pay back to the SO on aggregate at imbalance in both scenarios. However, SIP reduced this liability by £325m (75\% reduction) for AMV, and £193m (65\% reduction) for ZMV due to changes in SP. Similarly, generators pay back to the SO on aggregate at imbalance, but notably this liability is also higher (by £60m/£42m; SIP reduced this liability by 21\%/16\%) in the AMV/ZMV cases. SIP has two main effects on the system imbalance (and hence SP) which explain imbalance costs for both suppliers and generators being lower in the factual case. First, suppliers' and generators' imbalances are less strongly (positively) correlated with the overall system imbalance in the factual case. This means there are more occasions in which the supplier/generator has a `helpful' imbalance, settled at a favourable price. For example, SIP means suppliers are long when the system is short more often, and suppliers benefit from high imbalance payments. Second, when suppliers'/generators' imbalances are in the same direction as the overall system imbalance, SIP tends to reduce the overall system imbalance, meaning these imbalances are settled at a less extreme price. Table~\ref{tab:supplierGeneratorImbalance} summarises these differences.

\begin{table}[htbp]
  \centering
  \footnotesize
  \begin{threeparttable}
  \begin{tabularx}{\textwidth}{X c c c c c c}
\toprule
& \multicolumn{3}{c}{\textbf{Supplier}} &
  \multicolumn{3}{c}{\textbf{Generator}} \\
\cmidrule(lr){2-4}\cmidrule(lr){5-7}
\textbf{Metric (unit)}
& \textbf{Factual}
& \textbf{AMV}
& \textbf{ZMV}
& \textbf{Factual}
& \textbf{AMV}
& \textbf{ZMV} \\
\midrule
$S/G$ imbalance correlation with system imbalance
& 0.40 & 0.66 & 0.54
& 0.38 & 0.51 & 0.44 \\
$S/G$ imbalance in same direction as system imbalance (\%)
& 64.1 & 75.1 & 69.8
& 63.4 & 68.7 & 65.9 \\
SP when system and $S/G$ short (£/MWh)
& 168.89 & 189.23 & 180.74
& 164.21 & 173.21 & 172.04 \\
SP when system and $S/G$ long (£/MWh)
& 69.75 & 62.69 & 62.37
& 68.64 & 61.58 & 61.87 \\
\bottomrule
\end{tabularx}
  \end{threeparttable}
  \caption{Summary of the interaction between supplier and generator imbalance volumes and factual/counterfactual imbalance volumes and settlement prices.} 
  \label{tab:supplierGeneratorImbalance}
\end{table} 

Overall imbalance cashflows to Elexon, the settlement administrator, in the counterfactual are £106m/£179m higher, a 12\%/21\% increase, driven by higher imbalance costs to suppliers and generators. The change in total cashflow does not match the sum of the change in cashflows for suppliers and generators since the overall change in cashflow includes cashflow changes from other party types (such as NPTs or aggregators) which we do not include here. Changes to the imbalance cashflows of these other parties due to changes to the imbalance price explain the difference.

% Residual imbalance cashflows to the SO in the counterfactual scenarios are reallocated across the market BM units, using data which is not publicly available, hence we were unable to run this calculation. Since the generators are the primary owners of BM units, it is likely that generators lose out from NPT NIV chasing, through a combination of lower residual cashflow reallocation, lower BM revenue, and lower wholesale revenues (see Section~\ref{sec:discussionAndPolicyImplications}). We do not quantify the financial benefits to NPTs since this would require estimating the price at which the imbalance position was opened. However, it is safe to assume that this strategy is on aggregate profitable for NPTs, given the persistence of aggregate imbalance positions by NPTs.

\begin{figure}
    \centering
    \includegraphics[width=\linewidth]{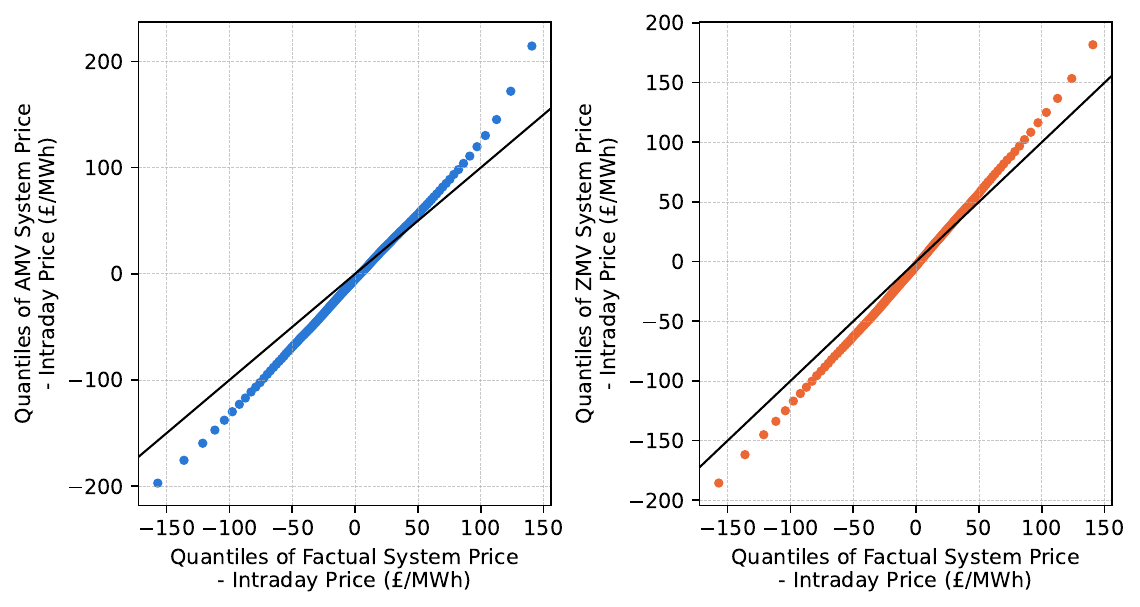}
    \caption{Q--Q plots displaying the quantiles of the difference between the counterfactual system price and the intraday price, and the factual system price and the intraday price in both AMV and ZMV scenarios. 1\textsuperscript{st}--99\textsuperscript{th} percentiles displayed.}
    \label{fig:imbalancePriceQqPlot}
\end{figure}

The financial effects on generators are likely asymmetric depending on the technology type. Smaller renewable generators, who are more prone to forecast errors and may be unable to self-balance their production, are more exposed to imbalance prices than larger firms with bigger portfolios and more conventional, controllable assets. To demonstrate this, Figure~\ref{fig:imbalancePriceQqPlot} plots the quantiles of $SP_{AMV/ZMV}-P_{ID}$ against $SP-P_{ID}$. $P_{ID}$ is an intraday reference price of the volume-weighted average price of all half-hourly contract trades on EPEX Spot in GB in the hour before gate closure. These plots aim to represent that SIP compresses real time scarcity margins in the BM. SIP in this time period will have some effect on the intraday price, generally in the opposite direction to the difference between the outturn SP and intraday price. As such, these plots \textit{underestimate} the extent to which SIP compresses scarcity margins. However, the plots reveal that the real time scarcity/oversupply premium is significantly larger without SIP. This indicates that this trading strategy closes arbitrage opportunities, and contributes to a more efficient intraday and balancing market. This suggests that SIP reduces imbalance price risk for other system users, with the largest benefits accruing to those parties most exposed to imbalance prices through forecast errors (namely suppliers and small renewable generators). Larger generating firms also lose out through reduced balancing revenues.

\section{Discussion and Policy Implications}
\label{sec:discussionAndPolicyImplications}

Ten years after the switch to single imbalance pricing, this paper marks the first research on how SIP affects the electricity system. Given the tolerance of the GB system towards opportunistic imbalancing and NPT market participation, this is an area of interest to electricity policymakers beyond GB considering the effects of switching to single imbalance pricing and allowing opportunistic imbalancing. Such relaxations could significantly alter market dynamics in, for example, Belgium and Germany. This research also provides the first major analysis of the role of NPTs in the GB system.

Our modelling indicates that SIP reduces consumer bills through a reduction in the pass-through of imbalance settlement costs to suppliers and BM costs incurred by NESO to consumers. Although we do not model the benefits of intraday trading with NPTs to suppliers, NPTs are net sellers in the intraday markets, which exerts downward pressure on prices. We would expect additional competition to encourage generators to offer closer to the competitive equilibrium price, likely contributing to lower wholesale costs, further reducing bills. We conclude that the incentives for SIP in the GB system align with the aims of the imbalance pricing modification, in reflecting the value of helpful imbalances to participants – NPTs, NESO, and consumers benefit from this practice.

While generators benefit from lower imbalance settlement costs, and renewable generators (who are most exposed to the imbalance price through forecast errors) in particular benefit from a less volatile imbalance price, the effects of SIP on generators on the whole are less clear-cut. First, reduced balancing volumes reduce generator BM revenues. Second, NPTs are net sellers in the intraday markets. Throughout, we assume the imbalances of all other parties are the same in the factual and the counterfactual scenarios. This does not entail that each NPT intraday sale displaces a generator sale: participants may otherwise self-balance their portfolios rather than using intraday trading, or the trade might not have been executed at all. However, further competition at intraday does reduce revenues to generators at the margin, at a benefit to purchasers who pay lower prices. In this sense, SIP represents a transfer from generators to NPTs, the SO and suppliers. 

Through changing the scheduled operations of units via ex-ante trading and the balancing volume requirement, SIP will affect carbon emissions. However, the effect of SIP on system emissions requires estimating which units respond to price movements in ex-ante markets, together with the trades corresponding to open positions for SIP. Neither of these inputs can be precisely calculated from publicly available data. As in Section~\ref{subsec:financialEffectsResults}, future research could extend the methodology of this paper to estimate the effects of SIP on ex-ante market prices and emissions. However, we now provide both quantitative and qualitative arguments to suggest that SIP reduces power system emissions. We note that the GB power system is subject to the UK emissions trading scheme, and as such a reduction in power system emissions may be offset by an increase elsewhere as out-of-merit producers sell on their allowances. However, reducing power system emissions is one of the key priorities for the UK government.

First, the UK government's preferred approach to calculating emissions changes to the GB power system, long-run marginal emissions factors \citep{UKPower2023_GHGValuation}, does not apply to this investigation, since we do not investigate a sustained change in demand. Instead, we calculate the balancing mechanism marginal emissions factor (MEF), as in \citep{lcpMef}. Essentially, this short-run statistic measures the short-run carbon cost of flexibility, and indicates how often NESO requires carbon intensive units to balance the system. A higher MEF implies a greater requirement to maintain carbon intensive units on hand for system balancing. In the factual case, the average MEF was 239 kgCO2e/MWh, compared to 279 and 272 kgCO2e/MWh in the AMV and ZMV cases. Lower balancing volumes reduce reliance on carbon intensive units for energy balancing.

Second, the further ahead of time the system is balanced, the larger the stack of plants which are able to change their output to help system balancing. This allows more efficient CCGTs (for example) to step in and increase their output, rather than relying on less efficient but more flexible peakers to respond to shortages. Similarly, running fewer plants closer to their full output improves emissions due to efficiency gains and a lower number of plant start-ups, and flatter output schedules reduce operational costs.

Third, although not a direct short-run carbon impact, SIP seems to both reduce the volatility of the settlement price (see Section~\ref{subsec:financialEffectsResults}) and allow smaller renewable generators to trade away their imbalances close to real time. This provides a long-run benefit to those most exposed to imbalance settlement prices, particularly variable renewable generators and suppliers. Reducing SP tail risk can help the business case for further renewable investment, as well as reducing hedging/overcaution costs from suppliers seeking to minimise their downside risk.

Our research relies on key assumptions and simplifications to the complexities of system balancing. Notably, we have neglected the interactions of SIP with transmission constraints, which is one of the key concerns raised regarding this trading strategy~\citep{nesoRnp}. This simplification does not affect the conclusions of our results on the benefits of SIP for supply and demand balancing. In practice, energy balancing must account for transmission constraints, and as such our merit order model is a conservative estimate of the reduction in balancing costs due to SIP, in keeping with our approach throughout. However, it is possible that changes to other units' schedules due to trades with NPTs could either help to resolve or aggravate a transmission constraint. For example, an NPT buying at intraday might induce a unit in an export constrained region to turn up, requiring the SO to reduce further generation in this region, while increasing generation elsewhere. Alternatively, an intraday trade in the other direction may alleviate the constraint, serving a dual purpose. Data currently available makes further analysis of this interaction challenging: counterparties of intraday trades are not available, constraint data in GB is scarce, and the correspondence between an executed trade and the change to a given unit's schedule is complicated by portfolio-level optimisation. Further research in this area, subject to improved data availability, would add to the research we present here.

Our analysis demonstrates the empirical benefits of SIP in a mature market with significant and growing renewables penetration for the first time. The proposal to realign the market closure deadline with gate closure (one hour in advance of delivery) \citep{remaUpdate} risks reducing SIP by NPTs. Given NPTs tend to open their positions within the final hour, reacting to updated forecasts and real time bid/offer acceptances by NESO, restricting SIP in this way would likely lead to lower activity, moving wholesale volumes into the BM, running counter to broader aims for the power market. Ultimately, a drop in SIP would harm consumers, suppliers, smaller generators, and NESO, with a likely gain for larger established generating firms. 

The results of this analysis cannot be applied directly to other European markets with similar structures, due to differences in information availability in particular. However, this research should provide a starting point for similar analyses in other systems, and demonstrates the possibility of overall system benefits to other countries considering permitting speculative imbalance positions.

\section*{Acknowledgements}

The authors thank David Mlynski, Nicholas Wood, Andrew Richards, and the wider team at Outlook Energy for helpful discussions and feedback on drafts of this manuscript. The authors would like to acknowledge that the DPhil studentship which made this research possible is funded by Outlook Energy through the University of Oxford.

\section*{Declaration of generative AI and AI-assisted technologies in the manuscript preparation process}

During the preparation of this work, the authors used Claude and ChatGPT in the implementation of the simulation code, and reviewing drafts of the manuscript. The authors reviewed and edited the output as needed and take full responsibility for the content of the published article. 

\bibliographystyle{elsarticle-harv} 
\bibliography{references}

\appendix
\section{Pseudocode for the counterfactual simulation}
\label{app:pseudocode}

This appendix sets out the counterfactual simulation described in Section~\ref{subsubsec:methods-BmSimulation}. The purpose is to provide a conceptual overview of the algorithms used in the simulation, and should be read in conjunction with the replication guide for the accompanying repository: \url{https://github.com/PSALOxford/niv-chasing-bm-analysis/tree/main}.

\paragraph{Notation}
Each settlement period is indexed by $t$. $NIV_t$ is the factual Net Imbalance Volume (NIV), in MWh, with $NIV_t>0$ meaning the system is short. The factual settlement stack $\mathcal{S}_t$ is the set of accepted balancing actions $a$. Each action has a BM Unit (BMU) $u_a$, a bid--offer pair number $k_a$ ($k_a>0$ for offers, $k_a<0$ for bids), a price $p_a$, a volume $q_a$ ($q_a>0$ for offers, $q_a<0$ for bids) and a System Operator (SO) flag $f_a$. $\mathcal{B}_t$ is the set of bid--offer pairs submitted for $t$. Each pair has a level $L_{u,k}$ in MW, which contributes $\tfrac12 L_{u,k}$ MWh over the half-hour SP. $\mathrm{PN}_u$, $\mathrm{MELS}_u$ and $\mathrm{MILS}_u$ are the Physical Notification, Maximum Export Limit and Maximum Import Limit, each integrated over the SP to give MWh. $\mathcal{P}$ is the set of BSC parties and $N\subseteq\mathcal{P}$ is the subset classified as non-physical traders (NPTs). For party $i$, $I_{i,t}$ is its energy imbalance volume and $C_{i,t}$ its imbalance charge, both from MR1B. Primes (e.g.\ $N'_t$, $\mathcal{S}'_t$, $P'_t$) denote counterfactual quantities.

\begin{algorithm}[htbp]
\footnotesize
\caption{Re-dispatch of the settlement stack to the counterfactual NIV}
\label{alg:redispatch}
\begin{algorithmic}[1]
\Function{Redispatch}{$\mathcal{S}, N, N', \mathcal{B}, H$}
  \State $\mathcal{S}' \gets \mathcal{S}$
  \If{$N>0$ \textbf{and} $N'>0$} \Comment{short in both cases}
    \If{$N' > N$}
      \State $\mathcal{S}' \gets \Call{Accept}{\mathcal{S}', N'-N, \mathcal{B}, H}$ \Comment{accept further offers}
    \Else
      \State $(\mathcal{S}', R) \gets \Call{Remove}{\mathcal{S}', N-N'}$ \Comment{remove marginal offers}
      \If{$N - R > N'$} \State $\mathcal{S}' \gets \Call{Accept}{\mathcal{S}', N' - (N-R), \mathcal{B}, H}$ \Comment{accept bids} \EndIf
    \EndIf
  \ElsIf{$N<0$ \textbf{and} $N'<0$} \Comment{long in both cases}
    \If{$N' < N$}
      \State $\mathcal{S}' \gets \Call{Accept}{\mathcal{S}', N'-N, \mathcal{B}, H}$ \Comment{accept further bids}
    \Else
      \State $(\mathcal{S}', R) \gets \Call{Remove}{\mathcal{S}', N-N'}$ \Comment{remove marginal bids}
      \If{$N - R < N'$} \State $\mathcal{S}' \gets \Call{Accept}{\mathcal{S}', N' - (N-R), \mathcal{B}, H}$ \Comment{accept offers} \EndIf
    \EndIf
  \Else \Comment{NIV changes sign}
    \State $(\mathcal{S}', R) \gets \Call{Remove}{\mathcal{S}', N}$ \Comment{remove energy actions in the factual direction}
    \State $\mathcal{S}' \gets \Call{Accept}{\mathcal{S}', N' - (N-R), \mathcal{B}, H}$ \Comment{accept actions in the opposite direction}
  \EndIf
  \State \Return $\mathcal{S}'$ sorted by price
\EndFunction
\end{algorithmic}
\end{algorithm}

\begin{algorithm}[htbp]
\footnotesize
\caption{Headroom available on each submitted bid--offer pair}
\label{alg:headroom}
\begin{algorithmic}[1]
\Function{Headroom}{$\mathcal{B}, \mathcal{S}, \mathrm{PN}, \mathrm{MELS}, \mathrm{MILS}$}
  \ForAll{BMUs $u$ in $\mathcal{B}$ and pairs $k$ of $u$, taken in order $k=1,2,\dots$ (offers) and $k=-1,-2,\dots$ (bids)}
    \State $A_{u,k} \gets \sum_{a\in\mathcal{S}: u_a=u,\,k_a=k} q_a$ \Comment{volume already accepted on the pair}
    \State $D_{u,k} \gets \mathrm{PN}_u + \tfrac12 \sum_{j \text{ before } k} L_{u,j}$ \Comment{energy delivered by earlier pairs}
    \If{$k>0$}
      \State $H_{u,k} \gets \max\!\big(\min(D_{u,k} + \tfrac12 L_{u,k},\ \mathrm{MELS}_u) - D_{u,k} - A_{u,k},\ 0\big)$
    \Else
      \State $H_{u,k} \gets \min\!\big(\max(D_{u,k} + \tfrac12 L_{u,k},\ \mathrm{MILS}_u) - D_{u,k} - A_{u,k},\ 0\big)$
    \EndIf
  \EndFor
  \State \Return $H$
\EndFunction
\end{algorithmic}
\end{algorithm}

\begin{algorithm}[htbp]
\footnotesize
\caption{Accepting additional balancing actions in merit order. The direction is $\sigma=\operatorname{sgn}(\Delta)$, with $\sigma=+1$ for offers and $\sigma=-1$ for bids.}
\label{alg:accept}
\begin{algorithmic}[1]
\Function{Accept}{$\mathcal{S}, \Delta, \mathcal{B}, H$} \Comment{add $|\Delta|$ MWh of new actions in direction $\sigma$}
  \State $\mathcal{M} \gets$ pairs in $\mathcal{B}$ of direction $\sigma$ in merit order (offers by ascending price; bids by descending price)
  \State $a^* \gets$ marginal energy action of direction $\sigma$ in $\mathcal{S}$: the most expensive unflagged offer ($\sigma=+1$) or cheapest unflagged bid ($\sigma=-1$) with non-zero de minimis-adjusted volume
  \If{no such $a^*$} \State $j_0 \gets 1$
  \ElsIf{$(u_{a^*},k_{a^*}) \in \mathcal{M}$} \State $j_0 \gets$ position of $(u_{a^*},k_{a^*})$ in $\mathcal{M}$
  \Else \State $j_0 \gets$ position of the pair in $\mathcal{M}$ priced nearest to $p_{a^*}$ that lies strictly beyond it in merit order
  \EndIf
  \State $r \gets |\Delta|$
  \For{$j = j_0, \dots, |\mathcal{M}|$} \Comment{continue up the merit order from the factual margin}
    \State $(u,k,\pi) \gets \mathcal{M}_j$; \quad $v \gets \min(|H_{u,k}|,\ r)$
    \If{$v>0$} \State $\mathcal{S} \gets \mathcal{S} \cup \{(u, k, p=\pi, q=\sigma v, f=\text{false})\}$; \quad $r \gets r - v$ \EndIf
    \If{$r \le 0$} \State \textbf{break} \EndIf
  \EndFor
  \State \Return $\mathcal{S}$
\EndFunction
\end{algorithmic}
\end{algorithm}

\begin{algorithm}[htbp]
\footnotesize
\caption{Removing accepted energy balancing actions from the margin inward. The direction is $\sigma=\operatorname{sgn}(\Delta)$, with $\sigma=+1$ for offers and $\sigma=-1$ for bids.}
\label{alg:remove}
\begin{algorithmic}[1]
\Function{Remove}{$\mathcal{S}, \Delta$} \Comment{remove up to $|\Delta|$ MWh of accepted actions in direction $\sigma$}
  \State $r \gets |\Delta|$
  \ForAll{$a \in \mathcal{S}$ with $f_a=\text{false}$ and $\operatorname{sgn}(q_a)=\sigma$, from the margin inward (offers by descending price; bids by ascending price)}
    \State $v \gets \min(|q_a|,\ r)$; \quad $r \gets r - v$
    \If{$v = |q_a|$} \State $\mathcal{S} \gets \mathcal{S}\setminus\{a\}$ \Else \State $q_a \gets q_a - \sigma v$ \EndIf
    \If{$r \le 0$} \State \textbf{break} \EndIf
  \EndFor
  \State \Return $\big(\mathcal{S},\ R = \sigma(|\Delta| - r)\big)$ \Comment{$R$: signed volume removed}
\EndFunction
\end{algorithmic}
\end{algorithm}

\begin{algorithm}[htbp]
\footnotesize
\caption{Counterfactual imbalance price (following BSC Section T, Annex T-1). $^\dagger$\,Most expensive in merit order: highest-priced buy actions if $N'>0$, lowest-priced sell actions if $N'<0$.}
\label{alg:price}
\begin{algorithmic}[1]
\Function{ImbalancePrice}{$\mathcal{S}', N', \mathrm{MIP}, \mathrm{BPA}, \mathrm{SPA}, \mathrm{TLM}$}
  \If{$\mathcal{S}' = \emptyset$} \State \Return $\mathrm{MIP}$ \EndIf
  \State Buy set $\mathcal{S}^{+} \gets \{a: q_a \ge 0\}$ by ascending price; sell set $\mathcal{S}^{-} \gets \{a: q_a<0\}$ by descending price
  \State \textbf{De minimis tagging:} $q_a \gets 0$ if $|q_a| < 0.1$ MWh and the total volume of the same BMU and pair is $< 0.1$ MWh
  \State \textbf{Arbitrage tagging:} net off each sell action against buy actions with price $\le$ its price
  \State \textbf{Flagging:} actions that are SO or CADL flagged are first-stage flagged. Buy actions priced above the most expensive first-stage-unflagged buy, and sell actions priced below the cheapest first-stage-unflagged sell, are second-stage flagged. If no action is first-stage unflagged, all actions are second-stage flagged.
  \State \textbf{NIV tagging:} net off buy and sell volumes, starting from the most expensive buy (if net short) or the cheapest sell (if net long)
  \State $\mathcal{R} \gets \mathcal{S}^{+}$ if $N'>0$ else $\mathcal{S}^{-}$, keeping only actions with non-zero remaining volume; \quad $\mathrm{PA} \gets \mathrm{BPA}$ if $N'>0$ else $\mathrm{SPA}$
  \If{$\sum_{a\in\mathcal{R}} q_a = 0$} \State \Return $\mathrm{MIP}$ \EndIf
  \If{any action in $\mathcal{R}$ is second-stage flagged}
    \State $\mathrm{RP} \gets$ volume-weighted price of the most expensive\textsuperscript{$\dagger$} 1 MWh (RPAR) of second-stage-unflagged actions in $\mathcal{R}$, or $\mathrm{MIP}$ if there are none
    \State Reprice second-stage-flagged actions at $\mathrm{RP}$ and re-rank $\mathcal{R}$
  \EndIf
  \State \textbf{PAR tagging:} keep only the most expensive\textsuperscript{$\dagger$} 1 MWh (PAR) of $\mathcal{R}$
  \State \Return $\dfrac{\sum_{a\in\mathcal{R}} \mathrm{TLM}_{u_a} q_a p_a}{\sum_{a\in\mathcal{R}} \mathrm{TLM}_{u_a} q_a} + \mathrm{PA}$ \Comment{$\mathrm{TLM}_u = 1$ if $u$ has no TLM}
\EndFunction
\end{algorithmic}
\end{algorithm}
\end{document}